\documentclass[conference]{IEEEtran}
\IEEEoverridecommandlockouts
\usepackage{cite}
\usepackage{amsmath,amssymb,amsfonts}
\usepackage{graphicx}
\usepackage{textcomp}
\usepackage{xcolor}
\usepackage{graphicx}
\usepackage{algorithmicx,algorithm,algpseudocode}
\usepackage{textcomp}
\usepackage{subcaption}

\usepackage{bm,amsthm}
\usepackage{listings}%
\usepackage{caption}
\usepackage{enumitem} \setlist[itemize]{leftmargin=*}
\usepackage{booktabs}
\usepackage{setspace,hyperref}
\usepackage{color}
\usepackage{etoolbox}
\usepackage{circuitikz}
\usepackage{tikz}
\usepackage[normalem]{ulem}
\usepackage{array}
\usepackage{multirow}
\usepackage{colortbl}
\usepackage{diagbox}
\usepackage{makecell}
\usepackage{boldline}
\usepackage{pgfkeys}

\usetikzlibrary{positioning}
\usetikzlibrary{shapes.symbols}
\usetikzlibrary{angles,quotes}

\newtheorem{lm}{Lemma}

\DeclareMathOperator{\diag}{diag}

\DeclareMathOperator{\vect}{vec}

\begin{document}

\title{Kronecker-structured Sparse Vector Recovery with Application to IRS-MIMO Channel Estimation}

\author{Yanbin He and Geethu Joseph\\
Signal Processing Systems Group, Delft University of Technology, The Netherlands\\Emails: \{y.he-1,g.joseph\}@tudelft.nl}

\maketitle

\begin{abstract}
We study the recovery of a sparse vector with a Kronecker structure from an underdetermined linear system with a Kronecker-structured dictionary. This problem arises in several applications, such as the channel estimation of an intelligent reflecting surface-aided wireless system. Existing work only exploits the Kronecker structure in support of the sparse vector and solves the entire linear system jointly with high complexity. Instead, we decompose the original sparse recovery problem into multiple independent subproblems and solve them individually. We obtain the sparse vector as the Kronecker product of individual solutions, retaining its Kronecker structure. Besides, the subproblems exhibit reduced effective measurement noise. Our simulations demonstrate that our method has superior estimation accuracy and runtime compared to the existing work. We attribute the low complexity to the reduced dimensionality of the subproblems and improved accuracy to the denoising effect of the decomposition step.
\end{abstract}

\begin{IEEEkeywords}
Basis expansion model, sparse Bayesian learning, singular value decomposition, angular sparsity
\end{IEEEkeywords}

\section{Introduction}
\label{sec:intro}

Compressed sensing (CS) has been extensively applied in diverse domains
and its success stems from the inherent redundancies of many real-world signals. With suitable bases, many signals can be represented by a small subset of bases and thus be \emph{sparsely} expressed \cite{kreutz2024dictionaries}. Further, CS is a powerful tool for estimating the unknown parameters of non-linear functions by exploiting sparsity using the basis expansion model (BEM) \cite{hastie2009elements}. Specifically, BEM expresses the parametric non-linear function as the product of a \emph{known overcomplete dictionary} of the basis functions and an \emph{unknown sparse coefficient vector}. Here, the basis functions are obtained by sampling the unknown parameter over its range. As only a few functions corresponding to the truth are activated, the coefficient vector is sparse and can be recovered by the CS techniques. We aim to study the sparse recovery problems arising in the context of BEM, applied to the intelligent reflecting surfaces (IRS)-aided wireless communication system for channel estimation.

Beyond the standard CS framework, BEM can exhibit additional structure, such as Kronecker-structured dictionary and Kronecker-structured support of the sparse vector, arising from image processing \cite{zhao2019exploiting} or wireless communications \cite{chang2021sparse,xu2022sparse,he2022structure}. This problem generally takes the following canonical~form,
\begin{equation}\label{eq.problem_basic}
    \bm y = \bm H \bm x+\bm n,
\end{equation}
where $\bm y \in \mathbb{C}^{\bar{M} \times 1}$ is the noisy measurement, $\bm H \in \mathbb{C}^{\bar{M} \times \bar{N}}$ is the dictionary with $\bar{M} < \bar{N}$, $\bm x \in \mathbb{C}^{\bar{N} \times 1}$ is the unknown sparse coefficient vector, and $\bm n$ is the measurement noise. The Kronecker-structured dictionary \cite{chang2021sparse,xu2022sparse,he2022structure,zhao2019exploiting,caiafa2013computing,duarte2011kronecker} can be represented as
\begin{equation}\label{eq.separable_dict}
    \bm H = \bm H_1 \otimes \bm H_2 \otimes \cdots \otimes \bm H_I = \otimes_{i=1}^I \bm H_i,  
\end{equation}
where $\bm H_i \in \mathbb{C}^{M_i \times N_i}$ with $\prod_{i=1}^I M_i = \bar{M}$ and $\prod_{i=1}^I N_i = \bar{N}$. Similarly, the support of $\bm x$ can be expressed as the Kronecker product of the $I$ support vectors of sizes $N_1,N_2,\ldots,N_I$. Here, the goal is to estimate $\bm x$ given the Kronecker-structured dictionary $\bm H$ and noisy measurement $\bm y$. 

The linear inversion problem~\eqref{eq.problem_basic} with Kronecker-structured dictionary~\eqref{eq.separable_dict} has been widely discussed \cite{chang2021sparse,xu2022sparse,he2022structure,zhao2019exploiting,caiafa2013computing,duarte2011kronecker}. One method is based on the orthogonal matching pursuit (OMP), which has low complexity but requires hand-tuning of the stopping criterion. Another approach employs the sparse Bayesian learning (SBL) framework called Kronecker SBL (KroSBL)~\cite{chang2021sparse,xu2022sparse,he2023bayesian,he2022structure}. KroSBL assumes a sparse-promoting prior on $\bm x$ to impose the Kronecker-structured support. It improves the recovery accuracy but at the cost of higher computational complexity. 

Further, in applications like intelligent reflecting surface (IRS)-aided wireless channel estimation, BEM can lead to a Kronecker-structured sparse $\bm x$, i.e., along with the support, the entries of $\bm x$ are also Kronecker-structured. For example, in IRS channel estimation~\cite{he2022structure,he2023bayesian}, BEM utilizes this angular sparsity by sampling pre-defined spatial angle grids for the angle-of-departure (AoD), angle-of-arrival (AoA), and the difference between the AoA and AoD. It constitutes an overcomplete dictionary with steering vectors of different combinations of the grid points. Then, the different combinations of AoDs and AoAs result in a Kronecker-structured dictionary and sparse coefficients. Mathematically, the channel estimation takes the form of \eqref{eq.problem_basic} and \eqref{eq.separable_dict}, and the sparse channel vector is given by
\begin{equation}\label{eq.vect_kro}
    \bm x = \otimes_{i=1}^I \bm x_i,
\end{equation}
where $\bm x_i \in \mathbb{C}^{N_i \times 1}$. However, existing algorithms~\cite{he2023bayesian,he2022structure} do not fully exploit prior knowledge of the Kronecker structure in \eqref{eq.vect_kro}. Therefore, we present an efficient method by enforcing~\eqref{eq.vect_kro} to solve~\eqref{eq.problem_basic}, with improved accuracy and runtime compared to~\cite{he2023bayesian,he2022structure}. Our contributions are two-fold:
\begin{itemize}
    \item \emph{Decomposition-based algorithm}: We solve the linear inversion problem by first decomposing measurements $\bm y$ into $I$ sub-vectors $\{\bm y_i\}_{i=1}^I$ to obtain each $\bm x_i$ separately, and reconstruct the solution to~\eqref{eq.problem_basic} as $\otimes_{i=1}^I \bm x_i$. 

    \item \emph{IRS-MIMO channel estimation}: Our method applied to IRS-MIMO channel estimation indicates that, due to the decomposition step, our algorithm achieves low complexity (with runtime reduced by two orders of magnitude) and implicitly reduces the noise level. This combination leads to better reconstruction performance than existing algorithms.
\end{itemize}
{Overall, our algorithm explicitly integrates~\eqref{eq.vect_kro} into the sparse vector inference process through the decomposition step that ensures low complexity, efficient implementation and enhanced reconstruction performance by enforcing prior knowledge and reducing noise, making it a practical choice for IRS-MIMO channel estimation.
}

\section{Kronecker-structured Sparse Recovery}
We study the Kronecker-structured sparse recovery problem of estimating the unknown vector $\bm x$ given by \eqref{eq.vect_kro} from the measurements $\bm y$ in \eqref{eq.problem_basic} and the Kronecker-structured measurement matrix $\bm H$ in \eqref{eq.separable_dict}. This section derives a decomposition-based algorithm to estimate the sparse vector $\bm x$. 

We start with the noiseless set of linear equations and then extend the algorithm to the noisy case. We begin with
\begin{equation}\label{eq.problem_basic_noiseless}
    \bm y = \bm H \bm x,
\end{equation}
with $\bm H = \otimes_{i=1}^I \bm H_i$, and $\bm x = \otimes_{i=1}^I \bm x_i$. 
We need the following lemma to devise a decomposition-based algorithm. 
\begin{lm}\label{lmm.kron_equations_separable}
    \cite[Lemma 4]{he2023bayesian} Consider a set of linear equations, $\left(\bm H_1 \otimes \bm H_2 \right)\left(\bm x_1\otimes \bm x_2 \right) = \bm y_1 \otimes \bm y_2$ with $\bm y_1,\bm y_2\neq \bm 0$. Solving for $\bm x_1\otimes\bm x_2$ from the linear equations is equivalent to solving $\bm H_1 \left(\alpha \bm x_1\right) = \bm y_1$ and $\bm H_2 \left(\alpha^{-1}\bm x_2\right) = \bm y_2 $, for any scalar $\alpha\neq 0$ accounting for the scaling ambiguity.
\end{lm}
Here, we can estimate individual vectors $\bm x_1$ and $\bm x_2$ only up to a scaling ambiguity denoted by $\alpha$. However, any $\alpha\neq 0$ leads to the same Kronecker product $\bm x_1\otimes \bm x_2 = (\alpha \bm x_1) \otimes (\alpha^{-1} \bm x_2)$. A trivial extension of the Lemma \ref{lmm.kron_equations_separable} to the Kronecker product of $I$ vectors immediately suggests that the noiseless problem~\eqref{eq.problem_basic_noiseless} can be decomposed into $I$ smaller sparse recovery problems, instead of jointly solving for all vectors $\{\bm x_i\}_{i=1}^I$. 

To elaborate, we first decompose $\bm y$ into low-dimensional vectors $\{\bm y_i\in\mathbb{C}^{M_i}\}_{i=1}^I$ such that $\bm y = \otimes_{i=1}^I \bm y_i$, to disentangle different $\bm x_i$'s into $I$ subproblems $\bm y_i=\bm H_i(\alpha_i\bm x_i)$, for some $\alpha_i$ with $\prod_{i=1}^I\alpha_i = 1$. As mentioned above, $\alpha_i$'s are not designable parameters but represent the inevitable scaling ambiguity in estimating $\bm x_i$'s separately. We find $\bm y_i$ using $(I-1)$ recursive rank-one decomposition. To circumvent the scaling ambiguity, we add a unit norm constraint to arrive at 
\begin{equation}\label{eq.decom_y_noiseless}
\bar{\bm y}_{i-1} = \bm y_{i}\otimes\bar{\bm y}_{i}\; \;\text{and}\;\;\|\bm {y}_{i}\| = 1,\;\;\forall i{\color{red}\in [I-1]},\\ 
\end{equation}
with $\bar{\bm y}_0=\bm y$, $\bar{\bm y}_{I-1}=\bm y_I$, and set $[I-1]:=\{1,2,\ldots,I-1\}$. Once $\{\bm y_i\}_{i=1}^I$ is obtained, we solve $I$ sparse recovery problems given by $\bm y_i = \bm H_i(\alpha_i\bm x_i)$ for $i\in [I]$ to obtain $\{\alpha_i\bm x_i\}_{i=1}^I$ and the final estimate of $\bm x = \otimes_{i=1}^I \alpha_i \bm x_i = \otimes_{i=1}^I \bm x_i$.

Extending to the noisy setting in \eqref{eq.problem_basic}, the decomposition step of $\bm y$ minimizes $\| \bm y - \otimes_{i=1}^I \bm y_i \|_2$, and we replace \eqref{eq.decom_y_noiseless} with $(I-1)$ rank-one approximations,
\begin{equation}\label{prob.bidecom}
(\bm y_{i},\bar{\bm y}_{i})=\underset{\substack{(\bm{z}\in\mathbb{C}^{M_i},\bar{\bm z}),\|\bm z\|_2=1}}{\arg\min}\|\bar{\bm y}_{i-1} - \bm z\otimes\bar{\bm z}\|_2,\forall i{\color{red}\in [I-1]},
\end{equation}
with $\bar{\bm y}_0=\bm y$ and $\bar{\bm y}_{I-1}=\bm y_I$. Then, \eqref{prob.bidecom} is solved recursively using singular value decomposition (SVD). {To see this, we examine $i=1$. We rearrange $\bm y$ as $\bm Y\in\mathbb{C}^{\bar{M}/M_1\times M_1}$ with $\vect(\bm Y)=\bm y$ and $\vect(\cdot)$ being vectorization. We note $\bm z\otimes \bar{\bm z}=\vect(\bar{\bm z}\bm z^\mathsf{T})$. Thus, cost of \eqref{prob.bidecom} is the same as the rank-one approximation $\|{\bm Y} - \bar{\bm z}\bm z^\mathsf{T}\|_\mathrm{F}$.} After SVDs, we have $I$ separate sets of noisy linear equations:
\begin{equation}\label{eq.deco_linear_inversion}
    \bm y_i = \bm H_i (\alpha_i\bm x_i)+\bm n_i,\; \; \forall i\in [I],
\end{equation}
where $\bm n_i$ is the noise term. We solve~\eqref{eq.deco_linear_inversion} using CS algorithms such as OMP or SBL. The resulting algorithm, denoted as the \underline{d}ecomposition-based \underline{S}parse \underline{R}ecovery (dSR), is in Algorithm~\ref{al.dKroSBL}. Here, the noise term $\bm n$ in \eqref{eq.problem_basic} needs not to admit a Kronecker structure. However, our method is effective in noisy scenarios and can even aid in denoising the signal, as the next.

\begin{algorithm}[t]
\caption{Decomposition-based sparse recovery}
\label{al.dKroSBL}
\begin{algorithmic}[1]
\Statex \textit {\bf Input:} Measurement $\bm y$, dictionaries $\{\bm H_i\}_{i=1}^I$

\For {$i=1,2,\ldots,I$}
\State Solve~\eqref{prob.bidecom} to obtain $\bm y_i$
\State Solve~\eqref{eq.deco_linear_inversion} for $\bm x_i$ using any CS algorithm
\EndFor
\Statex \textit {\bf Output:} Sparse vector $\bm x = \otimes_{i=1}^I \bm x_i$
\end{algorithmic}
\end{algorithm}

\subsection{Denoising Effect}\label{sec:denoise}
Denoising refers to the noise reduction in the measurements after decomposition, i.e., $\mathbb{E}\{\Vert\otimes_{i=1}^I \bm y_i-\bm H\bm x\Vert^2\}<\mathbb{E}\{\Vert\bm{n}\Vert^2\}$, where $\{\bm y_i\}_{i=1}^I$ are obtained after the decomposition step~\eqref{prob.bidecom}. To intuitively explain denoising, we reformulate the measurements $\bm{y}$ in \eqref{eq.problem_basic} as a rank-one matrix $\left( \otimes_{i=2}^I\bm H_i \bm x_i \right)\left( \bm H_1 \bm x_1 \right)^\mathsf{T}$ with noise using~\cite[Eq. (10)]{schacke2004kronecker},
\begin{equation}\label{eq.step_1}
    \bm Y = \left( \otimes_{i=2}^I\bm H_i \bm x_i \right)\left( \bm H_1 \bm x_1 \right)^\mathsf{T} + \bm N,
\end{equation}
where $\vect (\cdot)$ denotes vectorization, $\vect{(\bm Y)} = \bm y = \bm H \bm x+\bm n$, and $\vect{(\bm N)} = \bm n$.  The first step ($i=1$) of~\eqref{prob.bidecom}, yields $\bm y_1\otimes \bar{\bm y}_1$ as the estimate of $\alpha_1^{-1}(\otimes_{i=2}^I\bm H_i \bm x_i)$ and $\alpha_1\bm H_1 \bm x_1$, respectively, for some $\alpha_1$, approximating $\left( \otimes_{i=2}^I\bm H_i \bm x_i \right)\left( \bm H_1 \bm x_1 \right)^\mathsf{T}$ by $\bar{\bm y}_1\bm y_1^{\mathsf{T}}$. Therefore, it reduces noise by extracting the rank-one part of the measurement, filtering out higher-rank components that arise due to noise. So, the noise level reduces \cite{cai2018rate}, even after just the first step of~\eqref{prob.bidecom}, as characterized below.
\begin{lm}\label{lm.denoise}
    Suppose that the noise $\bm{n}$ in \eqref{eq.problem_basic} is zero-mean white Gaussian noise with variance $\sigma^2$. If $\bar{\bm y}_{1}$ and $\bm y_1$ are obtained from $\bm y$ using~\eqref{prob.bidecom}, then
    \begin{equation}
    \mathbb{E}\left\{\|\bm y_1\otimes \bar{\bm y}_1 - \bm H\bm x\|^2\right\} \approx \sigma^2( M_1-1+\bar{M}/M_1).
    \end{equation}
\end{lm}
\begin{proof}
    The result is a special case of perturbation analysis of low-rank tensor approximations in \cite{balda2016first} by setting the tensor order $R=2$ and $r$-ranks as $p_1 = p_2 = 1$ in \cite[Eq. (19)]{balda2016first}.
\end{proof}
Lemma~\ref{lm.denoise} states that after the first step of decomposition~\eqref{prob.bidecom}, the noise level $\mathbb{E}\{\Vert\bm{n}\Vert^2\}=\sigma^2\bar{M}$ reduces approximately as 
\begin{equation}
    \frac{\mathbb{E}\left\{\|\bm y_1\otimes \bar{\bm y}_1 - \bm H\bm x\|^2\right\}}{\mathbb{E}\{\Vert\bm{n}\Vert^2\}} \approx \frac{M_1-1}{\bar{M}}+\frac{1}{M_1}<1.
\end{equation}

We further corroborate the overall denoising  after $I-1$ steps of~\eqref{prob.bidecom} by comparing $\Vert\otimes_{i=1}^I \bm y_i-\bm H\bm x\Vert$ and $\Vert\bm y - \bm H\bm x\Vert$ via simulations in Table~\ref{tab.denoise}. Here, we generate the noiseless signal $\bm H\bm x$ with $I=3$ and $M_i=10$ for $i=1,2,3$ using~\eqref{eq.problem_basic}. We add zero-mean Gaussian noise $\bm n$ with different SNRs accordingly. Results show that the noise level is significantly reduced. Moreover, decomposition leads to $I$ low-dimensional sparse recovery problems, shortening  runtime, as discussed next.

\begin{table}[t]
\centering
\scriptsize
\caption{Demonstration of denoising with $M_i=10$ for $i=1,2,3$, using the original noisy signal $\bm y$, reconstructed signal $\otimes_{i=1}^3 {\bm y}_i$ after the decomposition step, and ground truth $\bm H \bm x$. }
\begin{tabular}{l|c|c|c|c|c|c}
\hline
SNR (dB) & 5      & 10    & 15    & 20    & 25    & 30    \\ \hline
$\|\bm y-\bm H \bm x\|_2^2$  & 34.661 & 9.553 & 2.853 & 0.972 & 0.304 & 0.090 \\ \hline
$\|\otimes_{i=1}^3 {\bm y}_i-\bm H \bm x\|_2^2$  & 0.980  & 0.272 & 0.080 & 0.027 & 0.009 & 0.002 \\\hline
\end{tabular}
\label{tab.denoise}
\end{table}

\subsection{Complexity Analysis}
The time and space complexities of Algorithm~\ref{al.dKroSBL} are $\mathcal{O}(M^{I+1}+IT_{\mathrm{CS}})$ and $\mathcal{O}(M^I+MN+S_{\mathrm{CS}})$, respectively, assuming $M_i=M$, $N_i=N$, and $I\ll M<N$. Here, $T_{\mathrm{CS}}$ and $S_{\mathrm{CS}}$ are the time and space complexities of the sparse recovery algorithm used with dSR. Also, all sparse recovery subproblems are independent of each other and can be solved in parallel. In that case, the time and space complexity dSR changes to $\mathcal{O}(M^{I+1}+T_{\mathrm{CS}})$ and $\mathcal{O}(M^I+IMN+IS_{\mathrm{CS}})$, respectively. We consider dSR with SBL for sparse recovery, namely decomposition-based SBL (dSBL), and compare the complexity in Table~\ref{tab.complexity} with alternating minimization-based KroSBL (AM-KroSBL) and SVD-KroSBL~\cite{he2023bayesian}.

\begin{table}
\centering
\scriptsize
\caption{Complexity of different schemes. $R_\mathrm{AM}$: the number of AM iterations. $R_\mathrm{EM}$: the number of EM iterations.}

\begin{tabular}{l|l|l}
\hline
Method     & Time Complexity & Space Complexity \\ \hline
AM-KroSBL  & $\mathcal{O}\big(R_\mathrm{EM}( R_\mathrm{AM} IN^I +(MN)^I )\big)$  & $\mathcal{O}((MN)^I)$         \\ \hline
SVD-KroSBL & $\mathcal{O}\big(R_\mathrm{EM}( N^{I+1} +(MN)^I )\big)$  & $\mathcal{O}((MN)^I)$         \\ \hline
dSBL       & $\mathcal{O}\big(R_\mathrm{EM}N^2MI + M^{I+1}\big)$   & $\mathcal{O}(M^I+MN+N^2)$       \\ \hline
\end{tabular}
\label{tab.complexity}
\end{table}

\subsection{Comparision with SVD-KroSBL}
The decomposition step of our algorithm resembles the optimization problem in the SVD-KroSBL algorithm~\cite{he2023bayesian}. However, they differ as KroSBL seeks a sparse vector with Kronecker-structured support without considering the Kronecker structure~\eqref{eq.vect_kro} of $\bm x$. It relies on Type-II learning to estimate the sparse vector $\bm x$, assuming that the measurement noise is Gaussian distributed, $\bm n\sim\mathcal{CN}(\bm 0, \sigma^2\bm I)$. KroSBL exploits the Kronecker-structured support of $\bm x$ by using a Gaussian prior distribution on $\bm x$ \cite{he2022structure} given by
 $   p(\bm x; \{\bm \gamma_i\}_{i=1}^I) = \mathcal{CN}\left(\bm 0, \diag(\otimes_{i=1}^I\bm \gamma_i)\right)$,
where $\{\bm \gamma_i\}_{i=1}^I$ are the unknown hyperparameters. Then, it solves for the maximum likelihood $\{\bm \gamma_i\}_{i=1}^I$ and uses $\{\bm \gamma_i\}_{i=1}^I$ to retrieve $\bm x$. In particular, SVD-KroSBL uses SVD in the $r$th iteration similar to our dSR as 
\begin{equation*}\label{prob.gammabidecom}
\bm\gamma_{i}^{(r)} =\underset{\substack{\bm\gamma_{i}: \|\bm \gamma_{i}\|_2=1,
\bar{\bm\gamma}_{i}\in\mathbb{R}^{N(I-i)}}}{\arg\min} \| \bar{\bm\gamma}_{i-1} - \bm \gamma_{i}\otimes\bar{\bm\gamma}_{i} \|_2, i\in [I-1],   
\end{equation*}
where $\bar{\bm \gamma}_{I-1}=\bm \gamma_I$ and $\bar{\bm \gamma}_0$ is from the previous iteration~\cite{he2023bayesian}. Although this is a decomposition step like~\eqref{prob.bidecom}, they differ in their specific approaches. First, adopting a prior using $\otimes_{i=1}^I\bm \gamma_i$ is to mimic the Kronecker-structured support of the sparse vector~\eqref{eq.vect_kro}, and not to decompose the original sparse recovery problem. 
Second, SVD-KroSBL is iterative where the sparse vector $\bm x$ is estimated \emph{jointly} using all $\{\bm \gamma_i\}_{i=1}^I$ as in~\cite[Eq. (18)]{he2022structure}, which leads to higher complexity, as shown in Table~\ref{tab.complexity}. Third, adopting $\bm\gamma = \otimes_{i=1}^I \bm \gamma_i$ does not ensure the Kronecker structure~\eqref{eq.vect_kro} in the sparse vector $\bm{x}$. Similarly, if assumption \eqref{eq.vect_kro} does not hold, our dSR framework fails. 

\begin{figure*}[t]
  \centering
  \begin{minipage}[t]{.32\linewidth}
      \centering
    \includegraphics[width=\textwidth]{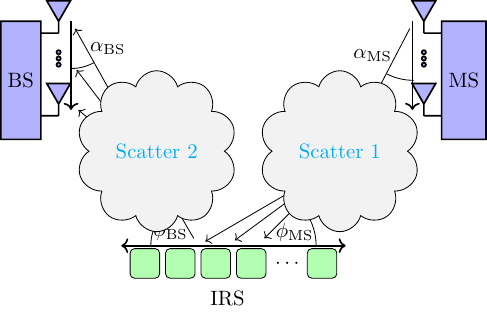}%
    \caption{An illustration of AoAs and AoDs in an IRS-aided uplink channel \cite{he2022structure}.}
    \label{fig:irs_sys}
  \end{minipage}\hfill
  \begin{minipage}[t]{.64\linewidth}
    \centering
    \includegraphics[width=.5\textwidth]{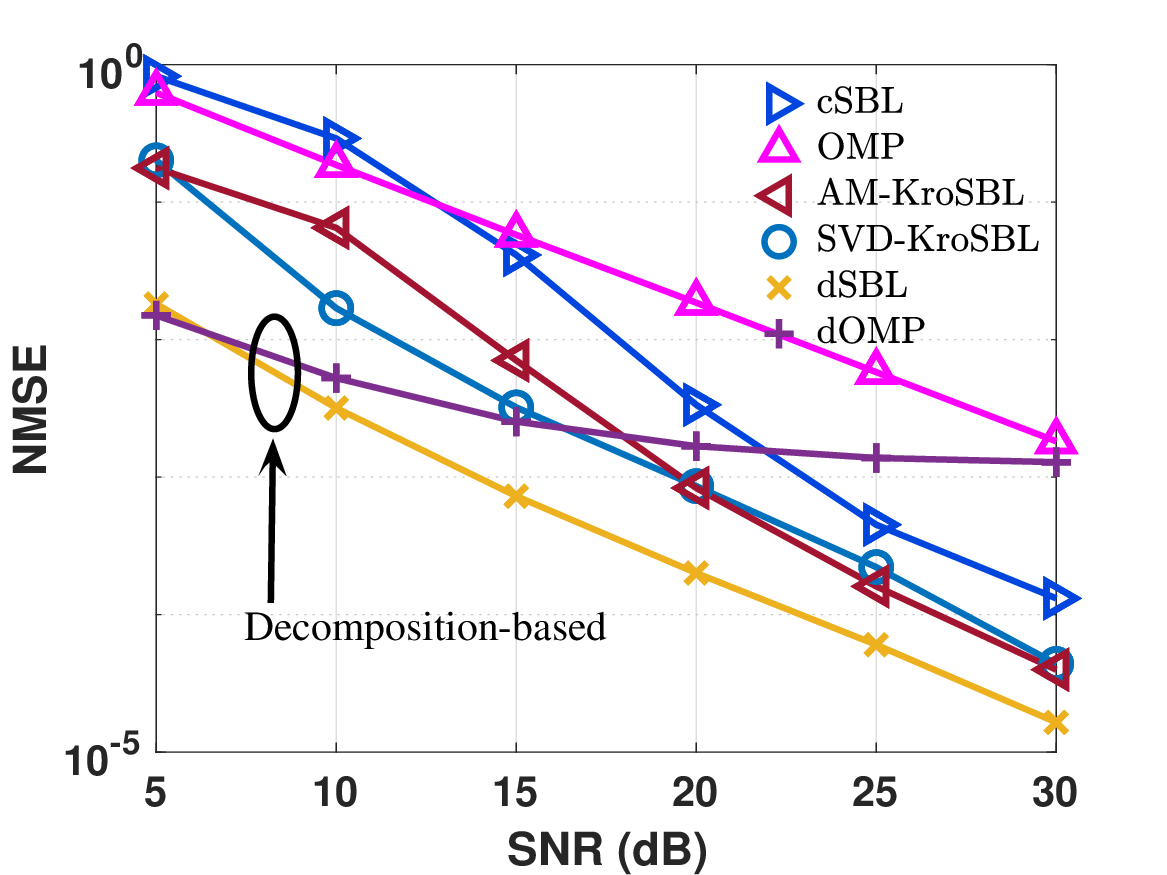}%
    \includegraphics[width=.5\textwidth]{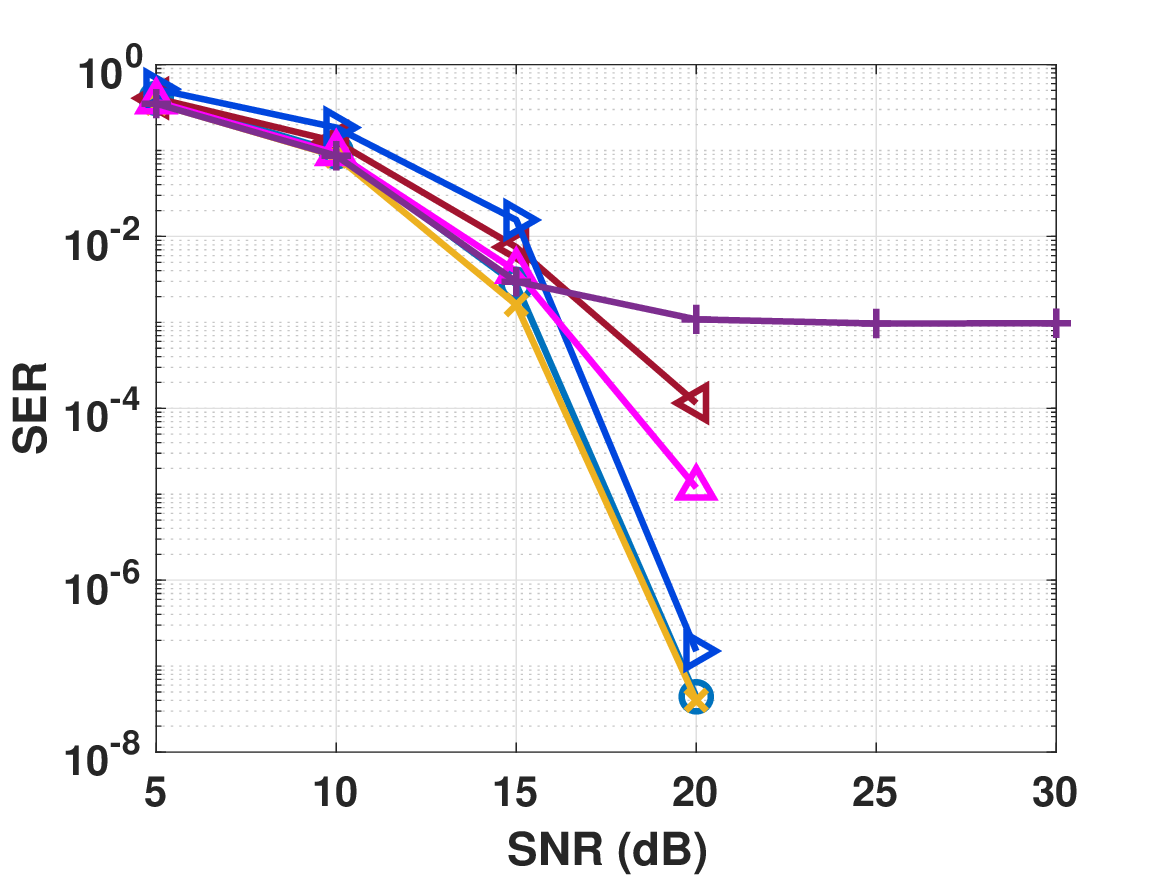}%
    \caption{NMSE and SER performance of different algorithms as functions of SNR with $K_{\mathrm{I}}=10$ and $K_{\mathrm{P}}=4$.}
    \label{fig.snr}
  \end{minipage}
\end{figure*}

\section{Cascaded Channel Estimation for a Prototypical IRS-aided System}
\label{sec:channelmodel}
In this section, we discuss the application of dSR to the problem of IRS-MIMO channel estimation. We consider an uplink narrowband millimeter-wave or terahertz band MIMO system with a $T$-antenna transmitter mobile station (MS) and an $R$-antenna receiver base station (BS), served by an $L$-element uniform linear array IRS. 

We start with the channel model in Fig.~\ref{fig:irs_sys}. The channel is the concatenation of the MS-IRS channel $\bm H_\mathrm{MS}$ and the IRS-BS channel $\bm H_\mathrm{BS}$. We adopt the geometric model~\cite{you2022structured,alkhateeb2014channel,wang2020compressed}~as
\begin{align}
\label{eq.channelmodel1}
\bm H_\mathrm{MS}&=\sum_{p=1}^{P_{\mathrm{MS}}}\sqrt{\frac{LT}{P_{\mathrm{MS}}}}\beta_{{\mathrm{MS}},p}\bm a_L(\phi_{{\mathrm{MS}},p})\bm a_{T}(\alpha_{{\mathrm{MS}}})^\mathsf{H}\\
\bm H_\mathrm{BS}&=\sum_{p=1}^{P_{\mathrm{BS}}}\sqrt{\frac{RL}{P_{\mathrm{BS}}}}\beta_{{\mathrm{BS}},p}\bm a_{R}(\alpha_{{\mathrm{BS}},p})\bm a_L(\phi_{{\mathrm{BS}}})^\mathsf{H},
\label{eq.channelmodel2}
\end{align}
where we define $\bm a_Q(\psi)\in\mathbb{C}^{Q\times 1}$ for an integer $Q$ and angle $\psi$ as $\bm a_Q(\psi) = 1/\sqrt{Q}[
1,e^{j\frac{2\pi d}{\lambda}\cos{\psi}},\cdots,e^{j \frac{2\pi d}{\lambda} (Q-1)  \cos{\psi}}]^\mathsf{T}$. Here, $d$ is the distance between two adjacent elements, and $\lambda$ is the wavelength. Also, $P_{\mathrm{MS}}$ and $P_{\mathrm{BS}}$ are the number of spread angles. We denote the $p$th AoA of the IRS, AoD of the MS, the $p$th AoA of the BS, and the AoD of the IRS as $\phi_{{\mathrm{MS}},p}$, $\alpha_{\mathrm{MS}}$, $\alpha_{{\mathrm{BS}},p}$, and $\phi_{{\mathrm{BS}}}$, respectively (see Fig.~\ref{fig:irs_sys}). 

For a given IRS configuration $\bm\theta \in \mathbb{C}^{L \times 1}$, the cascaded MS-IRS-BS channel is given by $\bm H_\mathrm{BS}\diag (\bm \theta)\bm H_\mathrm{MS}$. The $i$th entry of $\bm\theta$ represents the gain and phase shift by the reflection of the $i$th IRS element. Our objective is to estimate the cascaded channel $\bm H_\mathrm{BS}\diag (\bm \theta)\bm H_\mathrm{MS}$ for any given IRS configuration $\bm \theta$. 

We assume constant $\bm H_\mathrm{MS}$ and $\bm H_\mathrm{BS}$ over $K$ coherent slots. We send pilot data $\bm X \in \mathbb{C}^{T\times K_{\mathrm{P}}}$ spanning $K_{\mathrm{P}}$ time slots repetitively for $K_{\mathrm{I}}$ IRS configurations $\{\bm \theta_k\}_{k=1}^{K_{\mathrm{I}}}$ such that $K = K_{\mathrm{I}}K_{\mathrm{P}}$. Then, the received signal $\bm Y_k\in \mathbb{C}^{R \times K_{\mathrm{P}}}$ corresponding to the $k$th configuration $\bm \theta_k$~is $\bm Y_k=\bm H_\mathrm{BS}\diag (\bm \theta_k)\bm H_\mathrm{MS} \bm X+\bm W_k$ with $\bm W_k \in \mathbb{C}^{R\times K_{\mathrm{P}}}$ being the noise. 

To exploit the channel angular sparsity, we adopt a set of $N$ predefined grid angles $\{\psi_n\}_{n=1}^N$ with $\cos(\psi_n)= 2 n/N-1$~\cite{mao2022channel}. We collect steering vectors of the grid angles to formulate the BEM dictionaries
$$
\bm A_{Q} = \begin{bmatrix}
\bm a_{Q}(\psi_1) & \bm a_{Q}(\psi_2)&\ldots&\bm a_{Q}(\psi_N)
\end{bmatrix}\in\mathbb{C}^{Q\times N},
$$
for any integer $Q>0$ referring to the number of elements in the array. Then the BEMs of~\eqref{eq.channelmodel1} and~\eqref{eq.channelmodel2} are
\begin{equation}
\label{eq.channelmodelsparse2}
\bm H_\mathrm{BS}= \bm A_{R} \bm g_{\mathrm{R}} \bm g_{\mathrm{L,d}}^\mathsf{H} \bm A_{L}^\mathsf{H}\; \text{and}\;
\bm H_\mathrm{MS}= \bm A_{L}\bm g_{\mathrm{L,a}} \bm g_{\mathrm{T}}^\mathsf{H} \bm A_{T}^\mathsf{H},
\end{equation}
with $\bm g_{\mathrm{R}},\bm g_{\mathrm{L,d}},\bm g_{\mathrm{L,a}},\bm g_{\mathrm{T}}\in\mathbb{C}^{N\times 1}$ carrying the unknown channel state information, including AoAs/AoDs of the channel and path gains. Rearranging \eqref{eq.channelmodelsparse2} using $\bm Y_k$ and reorganizing the received signal $\{\bm Y_k\}_{k=1}^{K_I}$ (see \cite{he2022structure} for details), we arrive at 
\begin{equation}\label{eq.datamodelcs}
\tilde{\bm y} = (\bm \Phi_{\mathrm{L}} \otimes \bm \Phi_{\mathrm{T}} \otimes \bm \Phi_{\mathrm{R}})(\bm g_{\mathrm{L}} \otimes \bm g_{\mathrm{T}}^* \otimes \bm g_{\mathrm{R}}) + \tilde{\bm w} \in\mathbb{C}^{RK\times 1},
\end{equation}
where $\bm \Phi_{\mathrm{L}}\in\mathbb{C}^{K_{\mathrm{I}}\times N}$ is the first $N$ columns of $\bm \Theta^\mathsf{T}(\bm A_{L}^\mathsf{T} \odot \bm A_{L}^\mathsf{H})^\mathsf{T}$, $\bm \Phi_{\mathrm{T}}=\bm X^\mathsf{T} \bm A_{T}^*$, and $\bm \Phi_{\mathrm{R}}=\bm A_{R}$. Hence,~\eqref{eq.datamodelcs} transforms the channel estimation problem into a Kronecker-structured sparse recovery and we can apply Algorithm \ref{al.dKroSBL} to estimate the channel represented by $\bm g_{\mathrm{L}} \otimes \bm g_{\mathrm{T}}^* \otimes \bm g_{\mathrm{R}}$. Since there are three dictionaries, i.e., $\bm \Phi_{\mathrm{L}}$, $\bm \Phi_{\mathrm{T}}$, and $\bm \Phi_{\mathrm{R}}$, here $I=3$.

Extension to multiuser IRS-MIMO is also possible. The shared BS-IRS channel has the same sparse vector $\bm g_{\mathrm{R}}$ across users, but different user positions and pilot signals $\bm X$ lead to distinct $\bm g_{\mathrm{L}} \otimes \bm g_{\mathrm{T}}^*$ and $\bm \Phi_{\mathrm{L}} \otimes \bm \Phi_{\mathrm{T}}$. We modify~\eqref{eq.datamodelcs}~as
$$
\tilde{\bm y} = \left\{\sum_u(\bm \Phi_{\mathrm{L}} \otimes \bm \Phi_{\mathrm{T}}^{(u)})(\bm g_{\mathrm{L}}^{(u)} \otimes \bm g_{\mathrm{T}}^{*(u)}) \otimes \bm \Phi_{\mathrm{R}} \bm g_{\mathrm{R}}\right\} + \tilde{\bm w},
$$
where $u$ is the user index. By decomposing $\tilde{\bm y}\approx\bm y_1\otimes\bar{\bm y}_{1}$, we solve for $\bm g_{\mathrm{R}}$ and $\bm g_{\mathrm{L}}^{(u)} \otimes \bm g_{\mathrm{T}}^{*(u)}$ using linear models $\bar{\bm y}_{1} = \bm \Phi_{\mathrm{R}} \bm g_{\mathrm{R}}$ and $\bm y_1 = \sum_u(\bm \Phi_{\mathrm{L}} \otimes \bm \Phi_{\mathrm{T}}^{(u)})(\bm g_{\mathrm{L}}^{(u)} \otimes \bm g_{\mathrm{T}}^{*(u)})$, to obtain the channel estimates for each user. 

\section{Numerical Results: IRS-MIMO Channel Estimation}
\label{sec.numerical}

We apply dSR to the IRS-MIMO channel estimation problem using two versions of dSR: decomposition-based OMP (dOMP) and dSBL. For benchmarking, we implement classic SBL (cSBL) \cite{wipf2004sparse}, OMP \cite{cai2011orthogonal}, AM-KroSBL, and SVD-KroSBL~\cite{he2022structure}. We use $R=16$ BS antennas, $T=6$ MS antennas, $L=256$ IRS elements, $P_{\mathrm{BS}}=P_{\mathrm{MS}}=3$ spread angles, and spacing $d=\lambda/2$. The IRS configurations $\{\bm \theta_k\}_{k=1}^{K_{\mathrm{I}}}$ are selected uniformly from $\{-1/\sqrt{L}$ $,1/\sqrt{L}\}$ with $K_{\mathrm{I}}=10$. We send $K_{\mathrm{P}}=4$ pilot signals for each IRS configuration, resulting in $K= K_{\mathrm{I}}K_{\mathrm{P}}=40$ pilot signals. A predefined grid with $N=18$ angles is used in the BEM dictionaries, assuming uniform distribution of AoAs and AoDs. The channel gains $\{\beta_{{\mathrm{BS}},p}\}_{p=1}^{P_{\mathrm{BS}}}$ and $\{\beta_{{\mathrm{MS}},p}\}_{p=1}^{P_{\mathrm{MS}}}$ in \eqref{eq.channelmodel1} and \eqref{eq.channelmodel2} are drawn from the standard complex Gaussian distribution~\cite{lin2021channel}. The measurement noise is zero-mean white Gaussian. 

We use three metrics for comparison: normalized mean square error (NMSE) for channel estimation quality~\cite{lin2021channel,wang2020compressed,ardah2021trice}, symbol error rate (SER) for communication performance~\cite{he2022structure,he2023bayesian}, and runtime for complexity. Here, NMSE is 
$
    \frac{1}{K_{\mathrm{I}}}\sum_{k = 1}^{K_{\mathrm{I}}}\frac{\|\bm H_\mathrm{BS} \diag (\bm \theta_{k}) \bm H_\mathrm{MS} - \tilde{\bm H}_\mathrm{BS} \diag (\bm \theta_{k}) \tilde{\bm H}_\mathrm{MS}\|_\mathrm{F}^2}{\|\bm H_\mathrm{BS} \diag (\bm \theta_{k}) \bm H_\mathrm{MS}\|_\mathrm{F}^2},
$
where the channel estimate is $\tilde{\bm H}_\mathrm{BS}\diag (\bm \theta_{k}) \tilde{\bm H}_\mathrm{MS}$. We compute SER based on $10^6$ uncoded $8$-QAM symbols decoded using the channel estimate. 

Fig. \ref{fig.snr} shows that our approach outperforms the existing IRS channel estimation algorithms. Notably, dSBL has the best NMSE and SER. When SNR is low, dOMP outperforms all other methods except dSBL. As the SNR increases, dOMP continues to outperform OMP. Moreover, the under-sampling ratio is $KR/N^I=40\times  16/18^3\approx 10\%$, showing good performance in the lower measurement regime with limited pilot signals. This improved performance can be attributed to two factors: $i)$ we explicitly incorporate the Kronecker structure of $\bm g_{\mathrm{L}} \otimes \bm g_{\mathrm{T}}^* \otimes \bm g_{\mathrm{R}}$ in the channel estimation by the decomposition step in dSBL and dOMP, and $ii)$  the reduced noise perturbation in $\tilde{\bm y}$ after the decomposition (denoising). 

Furthermore, our approach offers a runtime that is at least two orders of magnitude shorter than existing algorithms. For example, when SNR = $20$~dB, the runtime in seconds is 12.595, 5.071, 2.904, 0.006, 0.001, and 0.165 for cSBL, AM-KroSBL, SVD-KroSBL, dSBL, dOMP, and OMP, respectively. Similar trends are observed across other SNR levels, underscoring the low complexity of our dSR framework.

\section{Conclusion}
We studied the sparse recovery problem with a Kronecker-structured dictionary and sparse vector, specifically in the context of channel estimation for IRS-MIMO systems. We first decomposed the problem into independent subproblems and solved them separately, exploiting the Kronecker structure. Our simulations showed that this approach improves the accuracy and runtime compared to the state-of-the-art methods, as the decomposition step leads to dimensionality reduction and denoising. Establishing theoretical guarantees and mathematically analyzing the denoising effect are exciting avenues for future research. 

\vfill\pagebreak
\bibliographystyle{ieeetr}
\bibliography{refs}

\end{document}